\documentclass{article}
\usepackage{graphicx}
\usepackage{amsmath}
\usepackage{hyperref}

\providecommand{\keywords}[1]
{
  \small	
  \textbf{\textit{Keywords---}} #1
}

\def\dbar{{\mathchar'26\mkern-12mu d}}

\title{Responding to objections made by Sorin Co\c sofre\c t \\ concerning kinetic molecular theory: \\ an illustration of how to teach physics students \\ to evaluate pseudoscientific work}
\author{Marc Meléndez Schofield}
\date{\today}

\begin{document}

\maketitle

\begin{abstract}
Sorin Co\c sofre\c t has argued that kinetic molecular theory does not correctly describe thermal equilibrium in ideal gases and has provided some examples that purportedly show that the standard laws imply consequences that contradict experimental results. This paper considers a few of the examples in detail and concludes that Co\c sofre\c t's analysis is erroneous. Furthermore, although it does not disprove any of the other claims made by Co\c sofre\c t in criticising thermodynamics and many other fields of physics, it does call his whole project into question, as Co\c sofre\c t exhibits a lack of relevant training, limited knowledge of key facts and previous research, and a disposition to favour his own intuitions above theoretical and experimental checks. The exercise below illustrates some of the problems commonly encountered in pseudoscientific approaches, and provides a few tips on how to teach aspiring scientists about how to evaluate such works for themselves.
\end{abstract}

\keywords{pseudoscience, educational physics, statistical mechanics, thermodynamics, kinetic molecular theory, molecular dynamics, computational physics}

\section{Introduction}

When I studied physics as an undergraduate, a man sometimes came to the campus with a banner and a megaphone and stood in front of the science buildings loudly accusing the mathematics and physics professors of being incompetent impostors. He claimed to have developed revolutionary theories that exposed the absurdities and errors of the theory of relativity, quantum mechanics, logic and set theory. His work was difficult to read and seemed to contain some mistakes, but while he complained about the lack of recognition from scientists, he showed little interest in discussing his manuscripts when confronted with perceived problems. I do not recall any professor or lecturer ever having responded to (or even mentioned) the man protesting outside. He claimed that scientific organisations and journals would consistently reject or disregard his ideas and the experiments he proposed. The apparent outright refusal to engage with him spurred some students into wondering whether he might in fact be a misunderstood genius. Most of us just assumed he did not know what he was talking about, but we had all heard the stories of Galileo's hardships, so every once in a while we na\"ively asked ourselves whether he might have been unfairly ignored.

Nowadays, thanks to the global reach afforded by social networks, so-called ``crackpot theories'' can quickly gain worldwide attention. Students not yet familiarised with the scientific endeavour might worry that researchers and institutions are trying to silence such dissident voices to protect the status quo. 

This article does not address the problem of demarcation \cite{Boudry2022}, that is, it does not aim to provide \textit{a priori} criteria for students to distinguish between science and pseudoscience \cite{Karaman2023,Erduran1995}. More often than not, pseudoscientific work gets rejected by professional journals not because it explicitly violates some point on a list of criteria, but because it does not meet the minimum standard of rigour required to substantiate the conclusions.

Science teachers have traditionally shown little interest in addressing pseudoscience topics \cite{Vlaardingerbroek2011}, but research suggests that refuting particular claims helps students avoid misconceptions and fosters critical thinking \cite{Manza2010,Kowalski2009}. Aspiring scientists can understand the problem with much pseudoscientific work and make up their minds for themselves by carefully examining one or two simple cases \cite{Schmaltz2014}. Here we illustrate the process by responding to a few simple claims made by Sorin Co\c sofre\c t regarding kinetic molecular theory.

Sorin Co\c sofre\c t has become relatively well known to physicists as the author of an impressive amount of spam email sent in the form of supposed scientific newsletters expressing criticism of a wide range of firmly established results in different fields, including mechanics, optics, thermodynamics, astrophysics, climate science, quantum mechanics, nutrition and immunology, among many others \cite{Cosofret-nl}. Scientific journals reject his submissions out of hand, according to his personal website at \url{www.pleistoros.com/en}. His work can also be found there, including a long list of previous newsletters which disparage the scientific community and treat many of the great achievements of their work with contempt. Most of the content lies far from my own field of expertise, but here I will concentrate on Co\c sofre\c t's objections to the concept of temperature in kinetic molecular theory and, especially, to the example he provides concerning a binary mixture of ideal gases \cite{Cosofret-temperature}, which he claims proves that the theory leads to predictions that contradict experiments.

I will end with some suggestions on how to teach aspiring scientists about common problems with many pseudoscientific presentations and why the scientific community and journals usually pay very little attention to them.

\section{The relation between temperature and kinetic energy}

According to the law of equipartition established by the standard classical theory of statistical mechanics \cite{Feynman2013,Tolman1979}, the average kinetic energy $\left<K\right>$ at equilibrium equals $k_B T/2$ per degree of freedom, with $k_B$ standing for Boltzmann's constant. The relation
\begin{equation}
    \left<K\right> =
    \frac{1}{2} m \left<v^2\right> = \frac{3}{2} k_B T
    \label{equipartition}
\end{equation}
provides the root mean square velocity $\bar{v}$ for an ideal gas molecule as a function of its mass $m$ and the temperature $T$.
\begin{equation}
    \bar{v} = \sqrt{\left<v^2\right>} = \sqrt{\frac{3 k_B T}{m}},
    \label{rmsv}
\end{equation}
or, if we use the molar mass $M$, $\bar{v} = \sqrt{3RT/M}$, with $R$ representing the universal gas constant.

\begin{figure}
    \centering
    \includegraphics[width=0.75\linewidth]{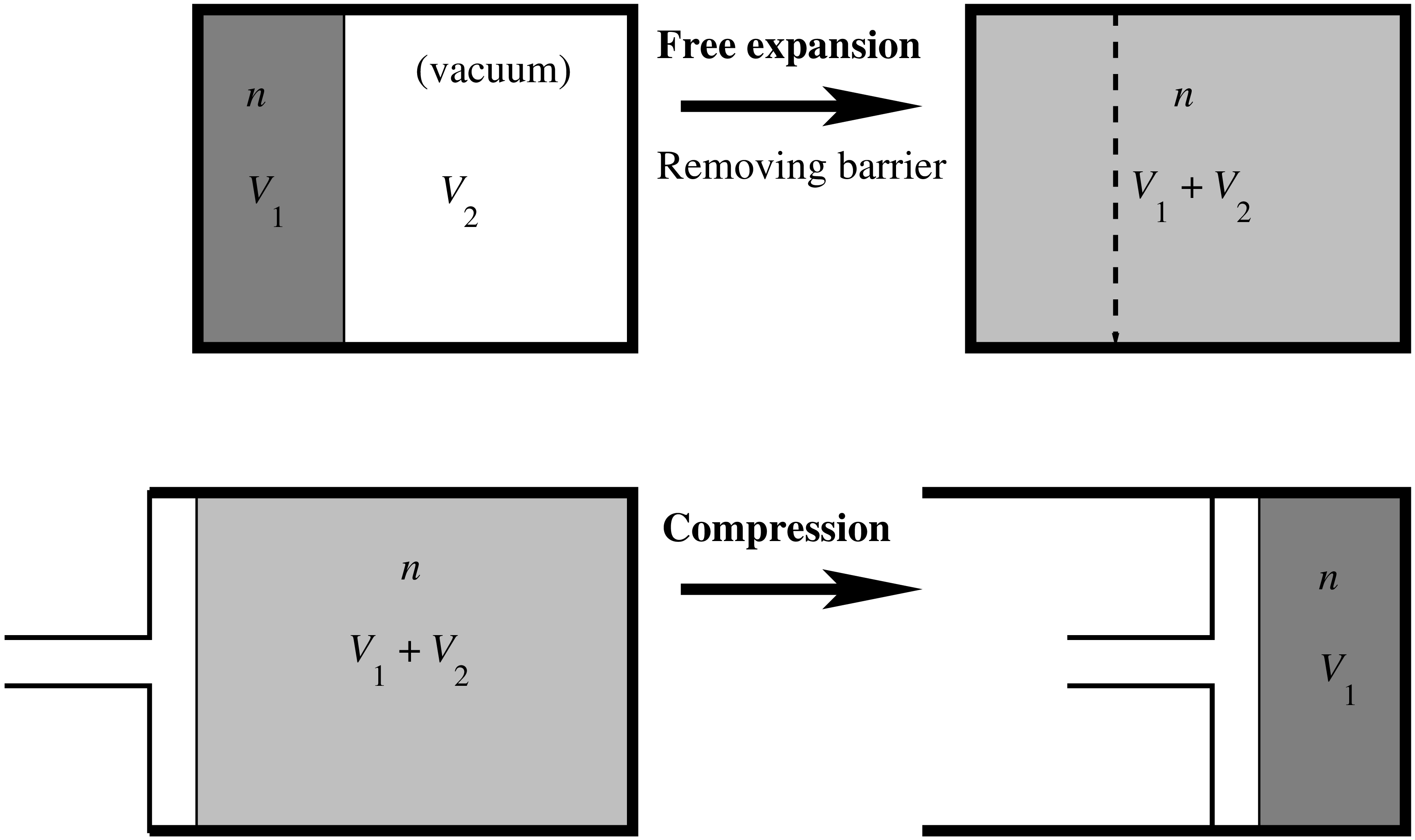}
    \caption{Standard thermodynamic processes of free expansion and compression of an ideal gas. \textit{Top}: $n$ moles of ideal gas contained in volume $V_1$ expand freely into an evacuated chamber of volume $V_2$ when the barrier between $V_1$ and $V_2$ is suddenly eliminated or breached. At equilibrium, the gas fills the total volume $V_1 + V_2$. \textit{Bottom}: A piston does work on the gas contained in a volume $V_1 + V_2$ slowly compressing it adiabatically to a volume $V_1$ and therefore increasing its temperature.}
    \label{expansion_compression}
\end{figure}

Co\c sofre\c t disputes this result and proposes two experiments to disprove this relation between temperature and kinetic energy. To the surprise of physicists and chemists, these concern the intensely studied and well-understood free expansion and slow adiabatic compression of a rarefied gas (Fig. \ref{expansion_compression}).

Let us first examine the free expansion, where the gas in a container is allowed to expand into an additional evacuated volume when the barrier between the original and new container is suddenly removed. Co\c sofre\c t correctly concludes that the theory for ideal gases implies that the temperature of the gas does not change when it expands freely into a larger volume. This agrees with the conclusion reached by Joule in 1845 when he carried out precisely this experiment \cite{Fermi1956}. In his own words, he stated that ``\textit{no change of temperature occurs when air is allowed to expand in such a manner as to not develop mechanical power}'' \cite{Joule1845}. In contrast, Co\c sofre\c t claims that the temperature will decrease.

It is unclear what to make of this objection, as Co\c sofre\c t does not present experimental results. Rather, he illustrates the phenomenon with a simplified diagram similar to Fig. \ref{expansion_compression} (\textit{top}) and a drawing of a thermometer with an imagined decrease of $5^\circ\mathrm{C}$. I can think of two ways to interpret the complaint.

On the one hand, Co\c sofre\c t could be referring to an actual small decrease in temperature measurable in experiments. This cooling has been studied and described in terms of more sophisticated models of gases that take inter-molecular forces into account \cite{Goussard1993}, but it makes no sense to complain that it is not predicted by the ideal gas model, which does not include inter-molecular potentials.

On the other hand, perhaps he is simply confusing a free expansion with an expansion against a piston, which does lead to cooling for an ideal gas. This interpretation becomes plausible when you realise that he describes compression of a gas with a piston (Fig. \ref{expansion_compression} \textit{bottom}) as reversing the other process, with an increase in temperature of $5^\circ\mathrm{C}$ in his illustration. He does not believe that the standard theory can explain the heating observed experimentally because a slowly moving piston cannot account for any significant changes in the temperature. He therefore says ``Can [kinetic molecular theory] or actual physics explain this increase of temperature for a compressed gas? The answer is definitely no''.

However, we can find solutions to this problem in college-level introductory thermodynamics textbooks \cite{Fermi1956b}. According to the first law of thermodynamics, the internal energy of the gas $U$ changes when energy flows into the system as work or heat, $dU = \dbar W + \dbar Q$. Here we concentrate on an adiabatic process, or $\dbar Q = 0$. The differential work equals $\dbar W = -PdV$, where $P$ is the pressure exerted by the piston, which equals the pressure of the gas if the compression proceeds slowly. Using the ideal gas equation of state $PV = nRT$, where $n$ stands for the number of moles, the differential work becomes $\dbar W = -nRTdV/V$. For monatomic ideal gases, such as helium or xenon, the internal energy corresponds to the total kinetic energy, $U = 3nRT/2$, using Eq. (\ref{equipartition}). The first law becomes
\begin{equation}
    \frac{3}{2}nR\ dT = -\frac{nRT}{V}\ dV. 
\end{equation}
If the process begins at temperature $T$, solving the differential equation reveals that the new temperature $T'$ after compression equals
\begin{equation}
    T' = T \left(\frac{V_1 + V_2}{V_1}\right)^{2/3}.
\end{equation}
Therefore, adiabatic compression of a gas of this kind to half its volume implies multiplying its temperature in Kelvin by a considerable factor of about $1.6$.

Although he presents no experimental data, Co\c sofre\c t accuses physicists of having ``no clue'' about the atomic processes taking place and merely ``talking from imagination'' \cite{Cosofret-temperature}. On the contrary, over the last century researchers have developed sophisticated models of microscopic phenomena which have achieved increasing agreement with experimental data (for a few examples, see Figs. 9--14, and 18 in Ref. \cite{Dorfman2014}). The discrepancies sometimes observed between experiments and theories indeed point to holes in our scientific knowledge, but remain irrelevant to Co\c sofre\c t's criticism as long as he does not propose theories that lead to even more precise predictions. Meanwhile, he shows little knowledge of the basics of thermodynamics theory or the experiments that led to discover its laws.

\section{Thermal equilibrium in a binary mixture of ideal gases}

Co\c sofre\c t claims that the analytical description of a binary mixture of ideal gases with different molecular weights cannot be found ``in any physics textbook or even in advanced treatises'' \cite{Cosofret-mixture}. Clearly, this claim was merely stated and not very carefully researched. Tolman's \textit{The Principles of Statistical Mechanics}, for example, covers precisely this kind of system \cite{Tolman1979b} in its Section \S35. Similarly, a quick internet search of ``kinetic theory for binary mixture of gases'' revealed as a top result a 1992 paper by Alves and Kremer \cite{Alves1992} which develops a more general theory for the behaviour of binary mixtures, which it then compares to experimental data. 

Co\c sofre\c t imagines a mixture of helium and xenon (molar masses: $M_\mathrm{He} = 4\ \mathrm{g/mol}$ and $M_\mathrm{Xe} = 131\ \mathrm{g/mol}$) at ambient temperature to illustrate his point. He estimates the velocity of the molecules from Eq. (\ref{rmsv}) and obtains $\bar{v}_\mathrm{He} = 1363\ \mathrm{m/s}$ and $\bar{v}_\mathrm{Xe} = 238\ \mathrm{m/s}$. Now, in a frontal elastic collision of molecules moving at these speeds, the helium atom would rebound having gained some speed, while the xenon atom would slow down, as we will show below with Eqs. (\ref{v_He}) and (\ref{v_Xe}). If the atoms crash head on into each other again, then the ``temperature difference'' increases even more. From this fact, Co\c sofre\c t concludes that kinetic molecular theory predicts that, in a mixture of helium and xenon initially at equilibrium, the helium would heat up and the xenon would cool down, which is obviously wrong. In his own words, he states that the change in speeds ``will lead to the existence of two different temperatures, and both will be different from the experimental one''.

Of course, the statistical-mechanical result applies to averages over many collisions and incidence angles and not a single frontal impact, but Co\c sofre\c t disregards this seemingly unimportant detail saying that although ``real collisions can take place under any angle of incidence,  the situation is more difficult to be treated mathematically, but the conclusions are the same'' \cite{Cosofret-mixture}.

We can easily show that the temperature of the \textit{mixture} remains constant due to the fact that elastic collisions conserve the total kinetic energy of the particles involved. Therefore, the total kinetic energy of the whole gas remains constant in spite of all the collisions. This implies that the average energy per particle $\left<K\right>$ must also be constant and, according to Eq. (\ref{equipartition}), the temperature will remain constant. However, Co\c sofre\c t wishes to consider the helium and xenon gases \textit{separately}. If the average kinetic energy per helium atom differs more and more from the value for xenon, then that would imply that the system tends to move away from equilibrium. This would lead to measurable effects (such as the helium becoming incandescent, for example) that we do not observe in experiments.

The first problem to deal with involves the interpretation of the system described. If it is supposed to represent a one-dimensional two-particle gas, then the velocities estimated above are wrong. This is because the law of equipartition applies to kinetic energies and velocities measured in the comoving frame of reference, that is to say, with respect to a coordinate system in which the gas remains at rest. In the frontal collision of atoms mentioned above, the total momentum does not vanish.
\begin{equation}
    m_\mathrm{He} v_\mathrm{He} + m_\mathrm{Xe} v_\mathrm{Xe} \neq 0.
\end{equation}
If we wish the correct equilibrium velocities, we should begin by measuring velocities in the reference frame with the system at rest. Define the momentum of helium as $p_\mathrm{He} = m_\mathrm{He} v_\mathrm{He}$ and similarly for xenon. In the comoving frame,
\begin{equation}
    p = p_\mathrm{He} = -p_\mathrm{Xe}.
\end{equation}
Therefore, at ``equilibrium'' the total kinetic energy would equal
\begin{equation}
    \frac{p^2}{2 m_\mathrm{He}} + \frac{p^2}{2 m_\mathrm{Xe}} = k_BT
\end{equation}
(note that in one dimension the average kinetic energy per particle equals $k_BT/2$). Solving for $p$,
\begin{equation}
    p = \sqrt{2 \frac{m_\mathrm{He} m_\mathrm{Xe}}{m_\mathrm{He} + m_\mathrm{Xe}} k_B T}.
\end{equation}
From here, we can derive the two velocities ($T = 298\ \mathrm{K}$),
\begin{align}
    v_\mathrm{He} & = \frac{p}{m_\mathrm{He}} = \sqrt{2\ \frac{M_\mathrm{Xe}}{M_\mathrm{He}} \frac{R T}{M_\mathrm{He} + M_\mathrm{Xe}}} = 1096\ \mathrm{m/s}, \label{v_He}\\
    v_\mathrm{Xe} & = -\frac{p}{m_\mathrm{Xe}} = -\sqrt{2\ \frac{M_\mathrm{He}}{M_\mathrm{Xe}} \frac{R T}{M_\mathrm{He} + M_\mathrm{Xe}}} = -33.5\ \mathrm{m/s}. \label{v_Xe}
\end{align}
Remember that we chose the average kinetic energy per particle in Eq. (6) to get the desired temperature of $T = 298\ \mathrm{K}$. Because elastic collisions conserve the total momentum and kinetic energy, the ``two-particle gas mixture'' will have the same kinetic temperature after the particles bump into each other, so the average kinetic energy and, by extension, the ``temperature'' of the two-atom gas would remain the same.

\begin{figure}
    \centering
    \includegraphics[width=0.75\linewidth]{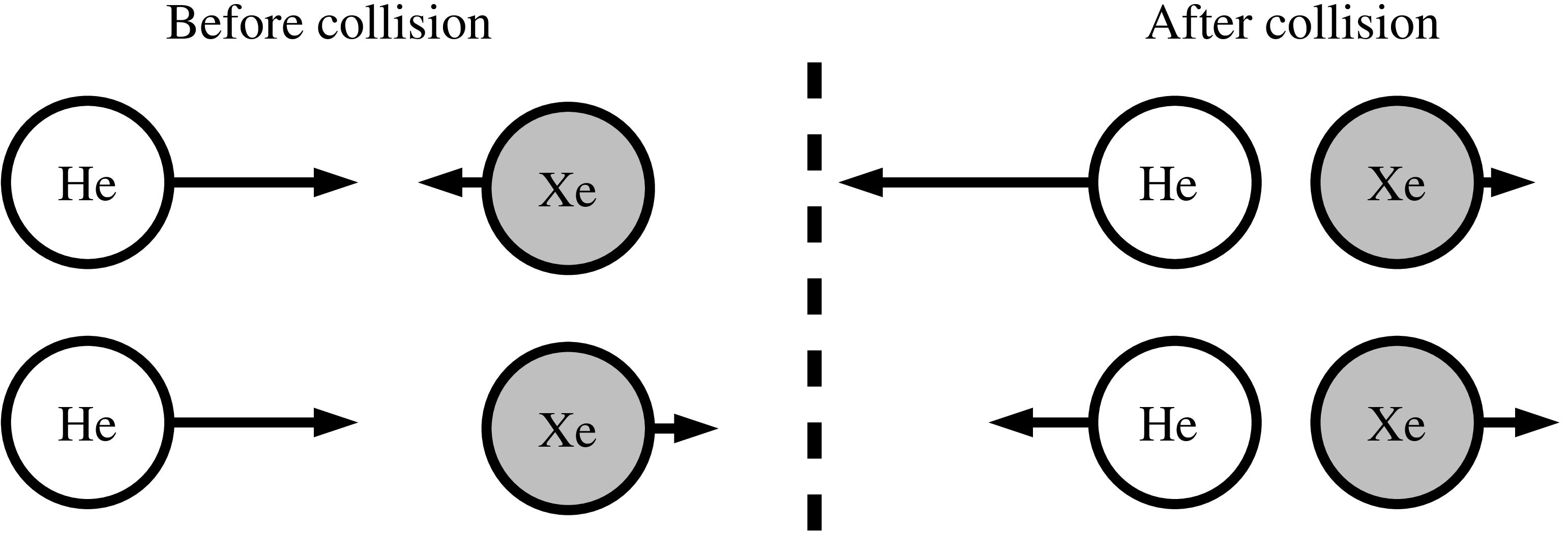}
    \caption{Two types of collision along the same axis for a helium and a xenon atom with speeds given by the average value $\bar{v}$ for each particle at thermal equilibrium. Because $\bar{v}_\mathrm{He} > \bar{v}_\mathrm{Xe}$, the He atom moves faster. At the top, they recoil after impact with the heavier xenon atom having slowed down and the lighter helium having accelerated. In the collision at the bottom, the lighter atom catches up with the heavier one and bumps into it from behind. After that, we find the opposite effect on the velocities, with the lighter atom moving slower than before and the heavier having sped up.}
    \label{collisions}
\end{figure}

Alternatively, we can interpret the helium and xenon atoms as two representative particles in a three-dimensional gas which happen to encounter each other straight on. In that case, we can use Eq. (\ref{rmsv}). For atoms moving in opposite directions, we can set $v_\mathrm{He} = 1363\ \mathrm{m/s}$ and $v_\mathrm{Xe} = -238\ \mathrm{m/s}$, but there would be more or less the same number of atoms of each element moving towards the right as to the left, and the centre of mass would remain at rest. Co\c sofre\c t's worry here comes from the fact that the collision he imagines accelerates the lighter particle and slows down the heavier one, so he concludes that according to the theory, the helium would heat up and the xenon would cool down even though they started at equilibrium. He does not realise that other collisions take place in which both particles move in the same direction. There we get the opposite effect, with the heavier particle accelerating and the lighter one slowing down (Fig. \ref{collisions}).

We will now verify that the \textit{average} speed remains unchanged after the collisions and that the helium and xenon in the binary mixture remain at the same equilibrium temperature. Elastic collisions conserve both the total momentum and total kinetic energy of the particles involved.
\begin{align}
    m_\mathrm{He} v_\mathrm{He} + m_\mathrm{Xe} v_\mathrm{Xe} & = m_\mathrm{He} v'_\mathrm{He} + m_\mathrm{Xe} v'_\mathrm{Xe}, \\
    \frac{1}{2} m_\mathrm{He} v^2_\mathrm{He} + \frac{1}{2} m_\mathrm{Xe} v^2_\mathrm{Xe} & = \frac{1}{2} m_\mathrm{He} {v'}^2_\mathrm{He} + \frac{1}{2} m_\mathrm{Xe} {v'}^2_\mathrm{Xe}.
\end{align}
The primes mark the velocities after impact, which we can isolate from the equations above.
\begin{align}
    v'_\mathrm{He} & = \frac{m_\mathrm{He} v_\mathrm{He} - m_\mathrm{Xe}(v_\mathrm{He} - 2 v_\mathrm{Xe})}{m_\mathrm{He} + m_\mathrm{Xe}}, \\
    v'_\mathrm{Xe} & = \frac{m_\mathrm{Xe} v_\mathrm{Xe} - m_\mathrm{He}(v_\mathrm{Xe} - 2 v_\mathrm{He})}{m_\mathrm{He} + m_\mathrm{Xe}},
\end{align}
Depending on the case considered (Fig. \ref{collisions}), we get different results. When $v_\mathrm{He} = 1363\ \mathrm{m/s}$ and $v_\mathrm{Xe} = -238\ \mathrm{m/s}$ (frontal collision), then
\begin{equation}
    v'_\mathrm{He} = -1744\ \mathrm{m/s},\ 
    v'_\mathrm{Xe} = -143\ \mathrm{m/s},
\end{equation}
but when the helium atom chases the xenon ($v_\mathrm{He} = 1363\ \mathrm{m/s},\ v_\mathrm{Xe} = +238\ \mathrm{m/s}$), then
\begin{equation}
    v'_\mathrm{He} = -820\ \mathrm{m/s},\ 
    v'_\mathrm{Xe} = 304\ \mathrm{m/s}.
\end{equation}

To calculate the average speed, we must note that the two cases from Fig. \ref{collisions} take place at different rates. In a unit time interval, a helium atom moving at speed $|v_\mathrm{He}|$ can reach a xenon atom moving in the opposite direction at speed $|v_\mathrm{Xe}|$ as long as it lies at a distance less than $|v_\mathrm{He}| + |v_\mathrm{Xe}|$. In the other case, the helium atom can catch up if the xenon lies less than $|v_\mathrm{He}| - |v_\mathrm{Xe}|$ away. Assuming a spatially homogeneous distribution of atoms, the fraction of head on collisions equals
\begin{equation}
    f = \frac{|v_\mathrm{He}| + |v_\mathrm{Xe}|}{|v_\mathrm{He}| + |v_\mathrm{Xe}| + |v_\mathrm{He}| - |v_\mathrm{Xe}|} = \frac{|v_\mathrm{He}| + |v_\mathrm{Xe}|}{2|v_\mathrm{He}|}.
\end{equation}
The weighted averages of the velocities after impact equal
\begin{align}
    |v'_\mathrm{He}| & = 1744\ f + 820\ (1 - f) = 1363\ \mathrm{m/s}, \\
    |v'_\mathrm{Xe}| & = 143\ f + 304\ (1 - f) = 238\ \mathrm{m/s},
\end{align}
which agree with initial the initial average velocities for a temperature of $T = 298\ \mathrm{K}$, according to Eq. (\ref{rmsv}).

To summarise this section, if we interpret Co\c sofre\c t's illustration as a one-dimensional system of two atoms, then he incorrectly applies the law of equipartition of energy because the velocities are not measured with respect to the centre of mass of the two atoms. If, instead, we consider the two atoms as representative members of a large collection, then he fails to account for the statistical character of the law. Careful examination of the collision of two particles with the speeds proposed shows that, when the averaging process is carried out correctly, the collisions do not change the temperature of the mixture.

\section{Numerical simulation of a binary mixture of gases\label{simulation}}

After the collisions described above, the atoms with new speeds can obviously collide again, changing their speeds repeatedly. What distribution of velocities emerges from all these collisions? Equilibrium statistical mechanics answers the question with the Maxwell-Boltzmann distribution \cite{Widom2002,Landau1980}. One can convince oneself of this result by going through the derivation carefully. A good introductory explanation can be found in chapter 40 of \textit{The Feynman Lectures on Physics} \cite{Feynman2013b}. However, Co\c sofre\c t does not address this theorem. Rather, his point is that molecular collisions would disrupt the equilibrium velocities. By calculating the typical velocities of two representative particles after impact, he argued that the velocities no longer corresponded to the right value for thermal equilibrium.

In the previous section we demonstrated that, 
quite the opposite, collisions among these representative particles indeed conserve the right \textit{average} velocity when we consider the two types of collision among particles with the specified speeds and the rates at which they take place.

Co\c sofre\c t stated without proof that his results will not change if we take into account the different directions along which particles can meet. Here we will refute his conclusions by presenting the results of a simplified numerical simulation of the binary mixture, where heavier particles will collide with lighter particles repeatedly, but the final distribution of velocities will agree with the initial equilibrium distribution.

\begin{figure}
    \centering
    \includegraphics[width=\linewidth]{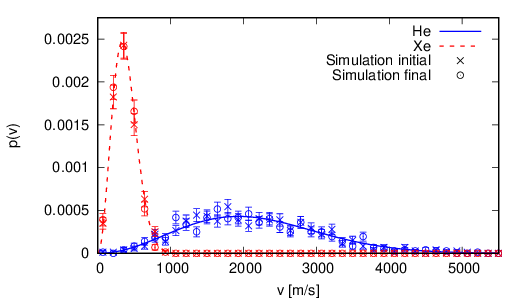}
    \caption{Theoretical Maxwell-Boltzmann distributions of velocities for He and Xe atoms (represented with solid and dashed lines, respectively) at $T = 298\ \mathrm{K}$, and velocity histograms in the numerical simulation of a binary mixture of He and Xe atoms. The initial speeds (marked with x) were chosen from the corresponding distributions, and the final speeds (circles) were calculated after each atom had undergone hundreds of collisions.}
    \label{Maxwell-Boltzmann}
\end{figure}

Let us set up a model of the binary mixture with a few hundreds of atoms of each type in a simulation box with periodic boundary conditions and let a computer calculate how the system evolves in time according to Newton's laws for a few hundred collision times. We give each atom an initial random direction and velocity in such a way that the speeds (that is, the norms of the velocity vectors) follow the theoretical Maxwell-Boltzmann distribution at temperature $T = 298\ \mathrm{K}$. After the simulation, we make histograms of the velocity distributions for the final state of the He and Xe atoms and compare them to the initial distributions and theoretical prediction (Fig. \ref{Maxwell-Boltzmann}). We do not need to worry about the relative sizes of He and Xe atoms or the interactions between them, because Co\c sofre\c t leaves these details out of his critique. The computational model that produced Fig. \ref{Maxwell-Boltzmann} started off with 1000 spheres of equal diameter $d$ placed into a $12\times12\times12\ d^3$ periodic box. Half of the particles were assigned masses and velocities for He atoms, while the others took values for Xe atoms.

As the system evolves in time, it samples states in the microcanonical or NVE ensemble (constant number of particles, volume and energy). One way to simulate this process would be to use the original event-driven molecular dynamics algorithm presented by Alder and Wainwright in 1959 \cite{Alder1959,Rapaport2004,Allen2017}. However, I found it more convenient to integrate the equations of motion over fixed time steps with a velocity Verlet code \cite{Rapaport2004b,Allen2017b} which I could easily adapt from other recent projects. I chose the Lennard-Jones 6--12 short-range interaction potential among atoms, as it has often been used to approximate the interaction among noble gas atoms. Nevertheless, the exact details of the interaction remain irrelevant in this context, as long as they lead to sufficiently accurate representations of elastic collisions. This requires a sufficiently small time step to keep the integration step error under control, as well as a low enough pressure and a high enough temperature to ensure that the atoms do not clump together in a condensed phase.

The mean free time $\tau$ between collisions for a sphere moving at velocity $\bar{v}$ in a volume $V$ filled with $N$ identical spheres is approximately
\begin{equation}
    \tau = \frac{4 V}{\pi d^2 \bar{v} N}.
\end{equation}
Carrying out the simulation for over $200\ \tau$, we see that the velocity distribution remains consistent with the theoretical Maxwell-Boltzmann curves at the end of the run (Fig. \ref{Maxwell-Boltzmann}). We chose the integration time step $\Delta t$ small enough to ensure that atom displacements during an integration step were  small compared to the sizes of the spherical atoms, so $\Delta t = 10^{-3}\ d/\bar{v}$.

Fig. \ref{Maxwell-Boltzmann} exhibits two features that illustrate the statistical nature of kinetic molecular theory. Firstly, one thousand atoms constitutes a very small system, when considered from the macroscopic point of view, so we expect fluctuations with respect to the theoretical curve for the gas in the thermodynamic limit. That explains why some points do not lie exactly on the curve. They would, however, approximate the curve more and more exactly if we increase the number of particles in the simulation or average over the distributions calculated at different times.

Secondly, we can use the histograms to estimate the frequency of any given velocity in the thermodynamic limit, but because we have a relatively small sample size ($N = 500$ for each species), we obtain non-negligible uncertainties (represented with error bars in Fig. \ref{Maxwell-Boltzmann}). These we can also reduce by increasing the number of particles or averaging over more snapshots.

The whole exercise presented above supports our confidence in kinetic molecular theory and shows conclusively the flawed character of Co\c sofre\c t's criticism of the link between kinetic energy and temperature.

\section{Conclusions and tips for teachers}

We can still admire Co\c sofre\c t's passionate pursuit of better scientific knowledge while recognising that his lack of relevant training causes serious mistakes and oversights in his reasoning.

In my experience, pseudoscientific proposals make scant use of comparisons between theories and empirical data or experimental checks, and Co\c sofre\c t's discussion shares the same shortcomings. His example with the free expansion of a gas does not even correctly describe the general facts that the theory is supposed to explain.

Nevertheless, this limited familiarity with the relevant available data does not by itself completely invalidate the critique of kinetic molecular theory considered here, which attempts to show that the accepted theories lead to conclusions that obviously contradict well known facts. The argument goes that collisions among the particles that make up an ideal gas would disrupt the equilibrium distribution of velocities. We have shown above that this is not the case, and that the mistaken conclusion arises from three errors, the first of which involves an incorrect interpretation of the meaning of the root mean square velocity (\ref{rmsv}), which should be measured with respect to a comoving reference frame. Secondly, Co\c sofre\c t misunderstands the law of equipartition for an ideal gas (\ref{equipartition}), which applies to statistical averages of the kinetic energy per particle. Finally, he fails to take into account the different angles at which collisions occur, and simply assumes that his conclusions for a head-on collision generalise to other incidence angles (and, I suppose, different velocities) without bothering to verify this intuition.

In his criticism, Co\c sofre\c t is not really engaging with a widely-accepted scientific theory, but rather arguing against his own misunderstandings and mistaken intuitions about the consequences of that theory. Given that he does not yet exhibit the basic knowledge and skills needed to correctly evaluate relatively straightforward scientific claims, as we have seen, his inadequate training calls his whole project into question. This does not make all his claims necessarily erroneous. Rather, it shows that the conclusions he reaches are unlikely to satisfy professional scientists who tend to hold themselves to stricter standards to justify their results.

Errors similar to those presented in this article would explain why professional journals would not accept Co\c sofre\c t's submissions. Editors typically reject pseudoscience not because they wish to defend accepted theories from the attack of alternative scientific theories, but because these submissions do not count as genuine scientific work. The problem does not lie with bold statements (although contradicting well established knowledge will certainly make readers skeptical). Instead, the fault comes from the quality of the reasoning put forth in favour of those statements.

When teaching science students about the difference between legitimate new theories and pseudoscientific claims, it can be useful to work through an example, concentrating on one or two concrete claims, as we have done here, to illustrate the ways in which basic misconceptions and errors in a few paragraphs can quickly lead professional scientists to disregard some presentations. The exercise should highlight the sloppy reasoning, lack of knowledge of basic facts, and reliance on vague intuitions, for example, and not just the fact that the conclusions contradict established results.

In my own experience, I have found that authors of pseudoscientific treatises often admit that their work might contain some mistakes, but that their production should be judged as a whole. This is not how we evaluate theories. Scientific knowledge should stand precisely on the details, on careful reasoning and agreement with experimental data. Accepted theories provide robust knowledge not because they agree with our intuitions, but due to the painstaking checking and rechecking, as even then errors sometimes creep in. Finding a serious misstep that affects the conclusions constitutes a valid reason for sending work back to the author so that it can be corrected.

Students should understand that it is not their job to refute every pseudoscientific pamphlet that they encounter, nor every incorrect statement in any single work. No one can possibly check every new scientific claim for themselves, even after filtering through editorial boards and peer review. Furthermore, although examining Co\c sofre\c t's criticism of kinetic molecular theory might not constitute exacting scientific work, it does involve some work nonetheless. It makes no sense to attempt to disprove every single dubious point in an essay that we suspect will not stand up to detailed scrutiny. It is better to pick a few well-understood points and concentrate on those. If nothing makes sense to you but the work does not appear to be mere nonsense, then it is all right to suspend judgment and leave the assessment to someone else.

As in the analysis presented above, the investigation can proceed in several steps of increasing difficulty, depending on the students' level. It can begin with searching the literature for information regarding the claims made by the author in question. Do they know the basic accepted facts? Are they acquainted with the relevant previous research and do they understand it correctly? If not, we will find it harder to take their theories seriously. Even more convincingly, perhaps a laboratory demonstration could be carried out (such as Joule's free expansion experiment in the case studied here) to challenge the author's statements.

Next, some introductory-level exercises might reveal additional problems. Here we showed that classical thermodynamics indeed predicts a variation of temperature due to the adiabatic compression of a gas mentioned before, in contrast to Co\c sofre\c t's claim that physics could not explain the increase in temperature.

A similar type of college-level physics calculation allowed us to apply the conservation of linear momentum and kinetic energy to determine the new average velocities after the collisions of representative atoms, and verify that the average kinetic energy remained unchanged by the collision.

A more advanced approach could incorporate relevant checks as part of a project. Here we showed that a molecular dynamics simulation outputs velocity distributions for the species in a binary mixture that coincide with the theoretical Maxwell-Boltzmann distribution after many collisions have taken place, refuting Co\c sofre\c t's claim that atomic collisions disrupt equilibrium.

Although the present response includes direct references to the source of Co\c sofre\c t's critique, pointing out the author by name will probably not be required or even desirable in the context of teaching. We do not want to encourage picking on the author because of a lack of relevant training in some field, nor is our intention to get the author to amend their ways. In my experience, conversations with the authors of pseudoscientific treatises rarely lead to anything but frustration. If this paper were brought to Co\c sofre\c t's attention, I doubt it would cause any major changes in his opinions.

Nothing in this paper is intended to make scientists immune from criticism, but that problem has little to do with the flimsy reasoning and dubious conclusions of many pseudoscientific theories. All too often, the authors of such works take legitimate criticism from experts regarding details of the publishing process or the state of a field of reasearch as an excuse to justify why their ideas have not gained acceptance. Realising that such theories cannot usually stand on their own merits helps students see that they are not alternative science that the ``establishment'' refuses to consider, and that training in a scientific discipline does not amount to joining a cult centred around faith in scientific consensus.

\end{document}